# ENDOR study of nitrogen hyperfine and quadrupole tensors in vanadyl porphyrins of heavy crude oil

I.N. Gracheva [1,*], M.R. Gafurov [1], G.V Mamin [1], T.B. Biktagirov [1], A.A. Rodionov [1], A.V. Galukhin [1], S.B. Orlinskii [1]

[1] Kazan Federal University, Kremlevskaya, 18, Kazan 420008, Russia

*E-mail: subirina@gmail.com

We report the observation of pulsed electron-nuclear double resonance (ENDOR) spectrum caused by interactions of the nitrogen nuclei $^{14}$N with the unpaired electron of the paramagnetic vanadyl complexes $VO^{2+}$ of vanadyl porphyrins in natural crude oil. We provide detailed experimental and theoretical characterization of the nitrogen hyperfine and quadrupole tensors.

## 1. Introduction

In recent years, numerous studies have been focused on characterization of trace metal complexes in crude oil [1, 2]. As it was postulated [3, 4], those are formed from biological compounds, so they can serve as biomarkers of oil reservoir formation [5]. These complexes are known to have negative effect on oil production process, so their chemical structure and spectroscopic properties are of great interest [1, 6, 7].

Since the most abundant metal complex in crude oil is vanadyl porphyrin, its chemical composition and electronic structure has been extensively studied by various analytical methods, including electron paramagnetic resonance (EPR) [8 - 11]. The focus of these works is often aimed toward possible contribution of porphyrins to the mechanisms of aggregation of oil fractions, especially the most heavy and metal-rich one, asphaltenes. There where various attempts of spectroscopic detection of noncovalent intermolecular interactions between vanadyl porphyrins and surrounding molecules [11 - 14], though it still remains an open question. It has been already shown that EPR is sensitive to structural distortions of the vanadyl porphyrin molecule [9, 15], so the EPR spectrum can potentially probe structural changes upon its binding with other molecules.

The vast majority of the works so far concerning the oil porphyrins are done either on the model systems or on the specially extracted oil fractions [16 - 18]. In the present paper, we report the possibility of detection of pulsed electron-nuclear double resonance (ENDOR) spectrum caused by interaction of the unpaired electron with the $^{14}$N nuclei of the porphyrin skeleton (see Fig. 1). We propose that $^{14}$N ENDOR spectra study can complement conventional EPR measurements in shedding light on structural perturbations of the complex upon intermolecular interactions. As it is pointed out in [19], specified thermal treatment of crude-oil containing species can lead to manifestation of interaction with the neighboring $^{14}$N nuclei even in the conventional EPR spectra. But, to the best of our knowledge, no $^{14}$N EPR, electron spin echo modulation (ESEEM) curves or ENDOR spectra were reported for the untreated, native crude oil samples [20 - 22].

## 2. Experimental Part

For our study we took a sample of heavy crude oil (room temperature density 930 kg/m$^3$, API gravity 20.7, viscosity 980 mPa·s) from the Ashalcha oilfield of Republic of Tatarstan (Russia). As it was reported earlier [9, 11], it manifests a typical EPR spectrum of vanadyl porphyrin and stable organic radical(s) shown in Fig. 2(a). The spectrum was measured using X-band (microwave frequency of about 9.5 GHz) Bruker Elexsys 680 spectrometer by means of field-swept two-pulse echo sequence $\pi/2 - \tau - \pi$. The length of $\pi$ pulse was 32 ns and the time delay $\tau = 240$ ns.

Powder EPR spectrum of vanadyl complex VO$^{2+}$ ($^{51}$V$^{4+}$, 3d$^1$, electron spin $S = 1/2$, nuclear spin $I = 7/2$) can be described by a spin Hamiltonian consisting of electron Zeeman and $^{51}$V nuclear hyperfine terms with $g$-tensor and hyperfine coupling $A$-tensor of axial symmetry. Spectrum consists of 16 "lines", 8 lines for parallel and 8 lines for perpendicular complex orientations, as is schematically shown in Fig. 2(a). Each pair of lines corresponds to particular projection of $I$. Frames of $g$-tensor and $A$-tensor coincide with the molecular frame as it is shown in Fig. 1. The principal values of $g$ and $A$ tensors are listed in Tab. 1.

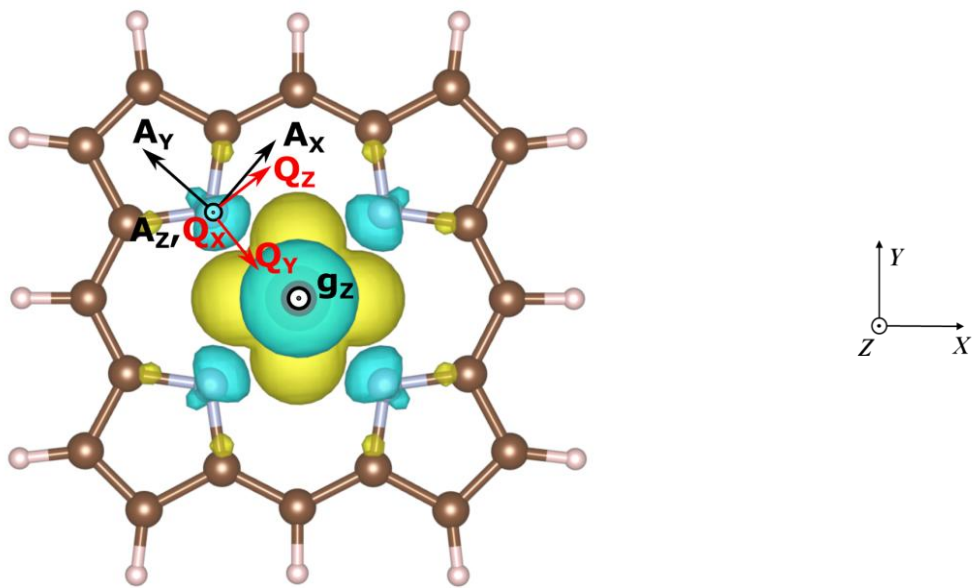

**Figure 1.** Schematic representation of a vanadyl porphyrin molecule considered in this paper (VO model). The orientations of nitrogen hyperfine ($A$) and quadrupole coupling ($Q$) tensors derived from DFT calculations are shown for a selected nuclei (consistent with Tab. 1). Isosurface illustrates the distribution of spin density. $X$-$Y$-$Z$ axes of the molecular frame are shown with the $z$-axis perpendicular to the porphyrin plane. As shown, the calculated $g_z$ is collinear with the molecular Z axis.

The interactions with $^{14}$N ($I = 1$, Larmor frequency $\nu_L = 1.06$ MHz at $B = 344$ mT) were probed with pulsed ENDOR measurements by using Mims sequence $\pi/2$-$\tau$-$\pi/2$-$T$-$\pi/2$ with an additional radiofrequency pulse $\pi_{RF} = 16$ μs inserted between the second and third microwave $\pi/2$ pulses at $T = 50$ K. The spectra are obtained at two different values of magnetic field, denoted as $B_1$ and $B_2$ in Fig. 2(a), in order to excite only those portions of the spins which are related to a certain molecular orientations. In our work we have chosen the values that correspond mainly to the $g_Z$ axis perpendicular ($B_1$) and parallel ($B_2$) to the direction of magnetic field for $m_I = 3/2$ transition due to the next reasons: (1) the sufficient echo amplitudes to obtain reasonable signal-to-noise ratio for the appropriate time; (2) absence of overlapping with the FR signal. As a result, we report observation of the ENDOR signals near the Larmor frequency of $^{14}$N, as displayed in Fig. 2(b, c).

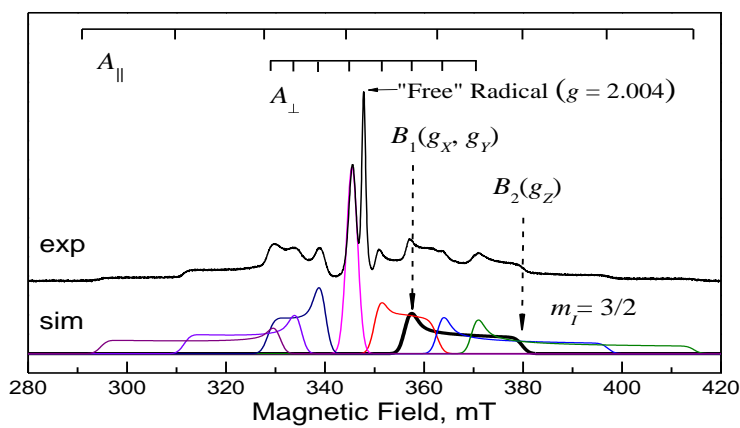

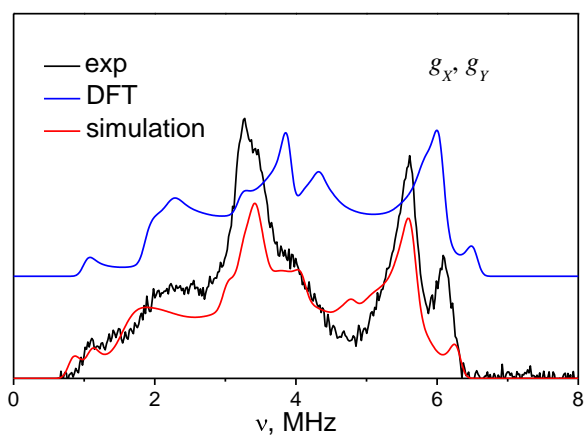

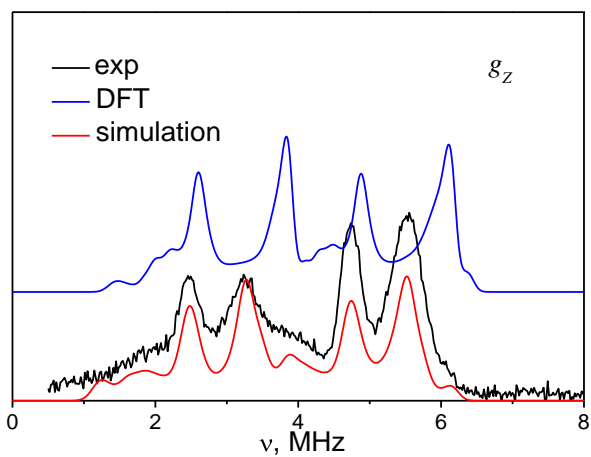

**Figure 2.** (a) X-band EPR spectrum of vanadyl porphyrin in crude oil sample in pulse mode at $T = 50$ K along with its simulation with the parameters from Tab. 1 (a). Magnetic fields $B_1$ and $B_2$ correspond to $g_z$ axis perpendicular and parallel to the direction of magnetic field ($m_I = 3/2$ transition). The signal at $g = 2.004$ is attributed to stable organic radicals. (b, c) $^{14}$N ENDOR spectra of vanadyl porphyrins (black), simulation (red) and calculated spectrum with parameters obtained by DFT calculations (blue): (b) – at magnetic field $B_1$, (c) – at magnetic field $B_2$. Corresponding parameters are listed in Tab. 1.

Table 1. Comparison between spin Hamiltonian parameters of vanadyl porphyrin complex in natural crude oil obtained from the simulation of the experimental EPR and $^{14}$N ENDOR spectra and calculated by DFT method ones for VO molecule[*]

|  | Simulation of experimental data | DFT (VO) |
|---|---|---|
| $g_X\ g_Y\ g_Z$ | (1.9845 1.9845 1.9640) | (1.9867 1.9867 1.9710) |
| $^{51}$V $A_X\ A_Y\ A_Z$ | \|156.9 156.9 470.8\| | (-157.0 -157.0 -463.9) |
| $^{14}$N $A_X\ A_Y\ A_Z$ | -6.5 -7.4 -7.8 | (-7.6 -7.8 -8.6) |
| $(\alpha\ \beta\ \gamma)_A$ | (90 0 0) | (-3 10 55) |
| $e^2Qq/h$ | 2.2 | 2.24 |
| $\eta$ | 0.50 | 0.22 |
| $(\alpha\ \beta\ \gamma)_Q$ | (30 90 180) | (40 90 -172) |

[*]Hyperfine couplings tensor components $A_{X,Y,Z}$ and quadrupole coupling constant $e^2Qq/h$ are in MHz. For $^{14}$N coupling parameters the values are averaged over the four pyrrole nitrogen nuclei. For VO models, the variance of both hyperfine and quadrupole coupling constants is within 0.1%. The Euler angles ($\alpha\ \beta\ \gamma$, in deg) are presented for a selected $^{14}$N nucleus (cf. Fig. 1) and specify Z-Y-Z rotation that transforms the molecular frame with the Z-axis being perpendicular to the porphyrin plane (see Fig. 1) to the frames where the hyperfine (A) and quadrupole (Q) tensors for $^{14}$N are diagonal. The coupling tensors for other three $^{14}$N nuclei are consistent with $C_{4v}$ symmetry of the molecule.

## 3. Discussion

The measured ENDOR spectra display the presence of both nuclear hyperfine and electric nuclear quadrupole couplings. In order to get some prior information about the values and orientations of the corresponding interaction tensors, we perform first-principles DFT calculations. Those are carried out in Orca program version 3.0 [23] using hybrid PBE0 exchange correlation functional, the 'Core prop' (CP(PPP)) basis set for vanadium, and EPR-II basis set for all other atoms. In sake of simplicity, we first consider a vanadyl porphyrin model which represents an isolated porphyrin skeleton with no side groups (we denote it as VO throughout the further discussion). Calculations were performed on HPC cluster for complex and demanding calculations of Kazan Federal University. The results of calculations are listed in Tab. 1. We notice that the g-tensor and $^{51}$V hyperfine coupling tensor are in a very good agreement with experiment. Next, we use the calculated $^{14}$N hyperfine tensor and quadrupole coupling tensor (represented by a coupling constant $e^2Qq/h$ and a rhombicity parameter $\eta$) as initial simulation parameters for the measured ENDOR spectra. One could anticipate a quite large error of the DFT calculated parameters, so during the further simulation we allow both the principal values and the tensor orientations to be varied within a consistent range. As a result, we end up with the coupling tensors that are given in Tab. 1 and Fig. 2(b, c).

The obtained experimental and calculated data are in a good correspondence with those for different vanadyls model systems [16 - 18, 22]. It is known that petroleum porphyrins exist in homologous manifolds of several structural classes and can manifest great structural diversity [24]. First of all that means different combinations of side groups. The distribution of electric field gradient and spin density in a vanadyl porphyrin molecule can be affected by the presence of substituting side groups via structural perturbations of the porphyrin skeleton. One can expect it to be especially the case if the groups are arranged asymmetrically, as for cycloalkane in VODPEP, thereby reducing the symmetry of pyrrole nitrogen atoms distribution around the vanadium ion. The consequent distribution of the coupling parameters can potentially increase the broadening of the spectrum, and this is probably why $^{14}$N ENDOR cannot be observed in certain (or in the most) oil samples. Following the discussion in Ref. [19], we propose that the possibility to detect $^{14}$N ENDOR depends on the relative presence of particular forms of vanadyl porphyrins in the sample. Further experimental studies and calculations are in progress.

## 4. Summary

To summarize, in this work we have detected for the first time pulsed $^{14}$N ENDOR spectrum of natural vanadyl porphyrins in the untreated heavy crude oil sample and provided detailed description of the corresponding nuclear hyperfine and quadrupole coupling tensors. We believe that $^{14}$N ENDOR can serve as an additional spectroscopic probe sensitive to structural perturbations of a porphyrin molecule, as supported by first-principles analysis.

## Acknowledgments

The authors devote this work to Dr. I.N. Kurkin (Kazan) on the occasion of his 75[th] anniversary. The work is financially supported by the Program of the competitive growth of Kazan Federal University among the World scientific centers "5-100" and RFBR grant No. 16-33-60085 mol_dk

The authors wish to acknowledge Kazan Federal University for the opportunity of performing first-principle DFT calculations on HPC cluster for complex and demanding calculations of KFU.

## References

1. Ventura G.T., Galla L., SiebertaC, Prytulaka J., Szatmari P, Hürlimann M, Halliday A.N. *Applied Geochemistry* **59,** 104 (2015)
2. Speight J.G. *The Chemistry and Technology of Petroleum, 5th edition* (CRC Press: Boca Raton, 2014), 942 p.
3. Treibs A. *Ann Chem* **510**, 42 (1934)
4. Treibs A. *Angew Chem* **49**, 682 (1936)
5. Baker E.W., Louda J.W. in *Biological Markers in the Sedimentary Record*, edited by R. B. John (Elsevier: Amsterdam, 1986), p. 125
6. Speight J.G. *Handbook of Petroleum Product Analysis, 2nd edition* (Wiley&Sons: Hoboken, 2015), 368 p.
7. Edelman I. S., Sokolov A. E., Zabluda V. N., Shubin A. A., Martyanov O. N. *Journ. Struct. Chem.* **57**, 382 (2016)
8. Trukhan S.N., Yudanov V.F., Gabrienko A.A., Subramani V., Kazarian S.G., Martyanov O.N. *Energy Fuels* **28**, 6315 (2014)
9. Biktagirov T.B., Gafurov M.R., Volodin M.A., Mamin G.V., Rodionov A.A., Izotov V.V., Vakhin A.V., Isakov D.R., Orlinskii S.B. *Energy Fuels* **28**, 6683 (2014)
10. Ramachandran V., van Tol J., McKenna A.M., Rodgers R.P., Marshall A.G., Dalal N.S. *Anal. Chem.* **87**, 2306 (2015)
11. Mamin G.V., Gafurov M.R., Yusupov .R.V., Gracheva I.N., Ganeeva Y.M., Yusupova T.N., Orlinskii S.B. *Energy Fuels* **30**, 6942 (2016)
12. Yin C.-X., Tan X., Müllen K., Bein T., Bräuchle C. *Energy Fuels* **22**, 2465 (2008)
13. Dechaine G.P., Gray M.R. *Energy Fuels* **24**, 2795 (2010)
14. Stoyanov S.R., Yin C.-X., Gray M.R., Stryker J.M., Gusarov S., Kovalenko A. *J Phys Chem B* **114**, 2180 (2010)
15. Espinosa M.P., Campero A., Salcebo R. *Inorg Chem* **40**, 4543 (2001)
16. Mulks C. F.; van Willigen H. *J. Phys. Chem.* **85**, 1220 (1981)
17. Gourier D., Delpoux O., Bonduelle A., Binet L., Ciofini I., Vezin H. *J Phys Chem B* **114**, 3714 (2010)
18. Smith II T.S., LoBrutto R., Pecoraro V.L. *Coordination Chemistry Reviews* **228**, 1 (2002)
19. Gilinskaya L.G. *Journ. Struct. Chem.* **49**, 245 (2008)

20. Atherton N.M., Fairhurst S.A., Hewson G.J. *Magn. Reson. Chem.* **25**, 829 (1987)
21. Galtsev V.E., Ametov I.M., Grinberg O.Ya. *Fuel* **74**, 670 (1995)
22. Fukui K, Ohya-Nishiguchi H, Kamada H. *J. Phys. Chem*. **97**, 11858 (1993)
23. Neese F. *Wiley Interdiscip Rev: Comp Mol Sci* **2**, 73 (2012)
24. Zhao X., Shi Q., Gray M.R., Xu C. *Scientific Reports* **4**, 5373 (2014)